\documentclass[twocolumn,superscriptaddress,
showpacs,preprintnumbers,amsmath,amssymb]{revtex4}

\usepackage{amssymb,amsmath,amsthm}
\usepackage{epsfig}
\newtheorem{Thm}{Theorem}

\theoremstyle{definition}

\newcommand{\bra}[1]{{\left\langle #1 \right|}}
\newcommand{\ket}[1]{{\left| #1 \right\rangle}}
\newcommand{\inn}[2]{{\left\langle #1 | #2 \right\rangle}}

\newcommand{\T}{\mbox{$\mathrm{tr}$}}

\begin{document}
\title{Bound on distributed entanglement}

\author{Jeong San Kim}\email{jkim@qis.ucalgary.ca}
\affiliation{
 Institute for Quantum Information Science,
 University of Calgary, Alberta T2N 1N4, Canada
}
\author{Soojoon Lee}\email{level@khu.ac.kr}
\affiliation{
 Department of Mathematics and Research Institute for Basic Sciences,
 Kyung Hee University, Seoul 130-701, Korea
}

\date{\today}

\begin{abstract}
Using the convex-roof extended negativity and the negativity of assistance
as quantifications of bipartite entanglement,
we consider the possible remotely-distributed entanglement.
For two pure states $\ket{\phi}_{AB}$ and $\ket{\psi}_{CD}$
on bipartite systems $AB$ and $CD$,
we first show that the possible amount of entanglement remotely distributed on the system $AC$
by joint measurement on the system $BD$
is not less than the product of two amounts of entanglement
for the states $\ket{\phi}_{AB}$ and $\ket{\psi}_{CD}$
in two-qubit and two-qutrit systems.
We also provide some sufficient conditions,
for which the result can be generalized
into higher-dimensional quantum systems.
\end{abstract}

\maketitle


\section{Introduction}

Quantum entanglement plays a crucial role in various kinds of quantum information tasks
such as quantum teleportation, dense coding, and quantum cryptography~\cite{tele, qkd1, qkd2}.
Whereas shared entanglement between different parties is generally consumed as a resource
in the tasks of quantum informational processing,
generating entangled states is usually expensive in practice.
Furthermore, due to the fragile nature of quantum entanglement,
entanglement in quantum states (and thus its non-classical ability for quantum information tasks)
may be gradually destroyed under the interaction with the environment,
which is known as the {\em decoherence}.
On this account,
it must be an important and necessary task
to provide an efficient way to create or distribute entanglement between desired parties,
and hence it is surely meaningful to quantify the possible amount of entanglement
distributed under given restricted conditions.

As a generalization of the entanglement swapping~\cite{ZZHE,ZHWZ,BVK},
the distribution of entanglement through quantum networks was characterized in qubit systems
by using the {\em concurrence}~\cite{ww} as a measure of bipartite entanglement:
If we have two states $\rho_{AB}$ and $\rho_{CD}$ in two-qubit systems $AB$ and $CD$ respectively,
the possible entanglement, which can be remotely distributed on the system $AC$
by arbitrary measurement on the system $BD$ and classical communication,
was shown to be always bounded above
by the product of two amounts of entanglement for $\rho_{AB}$ and $\rho_{CD}$~\cite{gs}.
Later, this bound was generalized to arbitrary qudit systems
by using the G-concurrence~\cite{g} as another bipartite entanglement measure.

However, for a bipartite pure state 
in a $d\otimes d$ quantum system,
the G-concurrence is defined as
the $d$th root of the determinant of its reduced density matrix 
with a proper normalization.
Although the G-concurrence has several good properties such as computability and multiplicativity~\cite{g},
it can only detect genuine $d$-dimensional entanglement
(that is, a bipartite pure state has a non-zero value of G-concurrence
if and only if its reduced density matrix has full rank),
and thus it has zero value for a lot of entangled states in two-qudit systems
whose reduced density matrices are not of full rank.
In other words, the G-concurrence cannot even give us a separability criterion,
which is one of requirements necessary for entanglement measures.

Another well-known quantification of bipartite entanglement is
the {\em negativity}~\cite{VidalW},
which is based on the {\em positive partial transposition} (PPT) criterion~\cite{Peres,Horodeckis1}.
The PPT condition is known to provide us with a separability criterion in the two-qubit system,
and also a necessary and sufficient condition for nondistillability
in $2\otimes d$ quantum systems~\cite{Horodecki1,DCLB}.
However, in higher-dimensional quantum systems rather than $2\otimes 2$ and $2\otimes 3$ systems,
there exist mixed entangled states with PPT,
the so-called PPT bound entangled states~\cite{Horodecki1,Horodeckis2},
that is, there exist entangled states whose negativity values are not positive.

In order to overcome the lack of separability criterion for mixed states,
the negativity can be modified by means of the {\em convex-roof extension},
which takes the minimal average of negativity values over all possible pure-state decompositions.
This modified negativity for mixed states
is called the {\em convex-roof extended negativity} (CREN)~\cite{LCOK},
and it gives a perfect discrimination between PPT bound entangled states and separable states
in any bipartite quantum system.

Here, we provide a new bound for remotely-distributed entanglement (RDE) using CREN
and its dual quantity, the {\em negativity of assistance} (NoA)~\cite{kds}.
Given a pair of pure states $\ket{\phi}_{AB}$ and $\ket{\psi}_{CD}$
in two bipartite systems $AB$ and $CD$ respectively,
we first show that the possible RDE on the system $AC$ by joint measurement on the system $BD$
is not less than the product of two amounts of CREN
for $\ket{\phi}_{AB}$ and $\ket{\psi}_{CD}$ in two-qubit and two-qutrit systems.
For $d\otimes d$ quantum systems ($d\ge 4$),
we provide some sufficient conditions of $\ket{\phi}_{AB}$ and $\ket{\psi}_{CD}$,
for which the result in lower-dimensional systems can be generalized.

This paper is organized as follows.
In Section~\ref{Subsec: definition},
we recall the definition of negativity, CREN, and NoA.
In Section~\ref{Subsec: lbound},
we derive an analytic lower bound of possible RDE for arbitrary dimensional quantum systems.
In Section~\ref{Subsec: qubits and qutrits},
we establish the inequality for the lower bound of the possible RDE
in low-dimensional quantum systems with respect to CREN and NoA.
In Section~\ref{Subsec: qudits},
we provide some sufficient conditions,
for which the result in low-dimensional systems can be generalized into arbitrary dimensional systems,
and we summarize our results in Section~\ref{Sec: Conclusions}.


\section{Convex-roof Extended Negativity and Negativity of Assistance}
\label{Sec: definition}
In this section,
we recall the definitions of CREN and NoA for bipartite quantum states.
We also provide an analytic lower bound of the distributed entanglement of $\rho_{AC}$,
which can be remotely prepared from two bipartite states $\ket{\phi}_{AB}$ and $\ket{\psi}_{CD}$
by a joint measurement in the system $BD$.

\subsection{Definitions}
\label{Subsec: definition}
 For a bipartite state $\rho_{AB}$ in a $d_A\otimes d_B$
quantum system, its negativity $\mathcal{N}(\rho_{AB})$
is defined as
\begin{equation}
\mathcal{N}(\rho_{AB})=\frac{\left\|\rho_{AB}^{T_B}\right\|_1-1}{d-1},
\label{negativity}
\end{equation}
where $\left\|\cdot\right\|_1$ is the trace norm, $d$=$\min\{d_A, d_B\}$,
and $\rho_{AB}^{T_B}$ is the partial transposition of $\rho_{AB}$~\cite{VidalW}.
For the case when $\rho_{AB}$ is a pure state $\ket{\psi}_{AB}$
with the following Schmidt decomposition,
\begin{equation}
\ket{\psi}_{AB}=\sum_{i=0}^{d-1} \sqrt{\lambda_{i}}\ket{ii},~~\lambda_{i} \geq 0,~~\sum_{i=0}^{d-1}\lambda_{i}=1,
\label{Schmidt}
\end{equation}
(without loss of generality, we may assume that the Schmidt basis is taken to be the standard basis),
its negativity can also be expressed in terms of its Schmidt coefficients, that is,
\begin{equation}
\mathcal{N}(\ket{\psi}_{AB})
=\frac{2}{d-1}\sum_{i<j}\sqrt{\lambda_{i}\lambda_{j}}.
\label{pure negativity}
\end{equation}

In order to compensate for the lack of separability criterion in the negativity,
CREN for a mixed state $\rho_{AB}$
is defined as
\begin{equation}
\mathcal{N}_c(\rho_{AB})= \min \sum_k p_k
\mathcal{N} (\ket{\psi_k}_{AB}), \label{CREN}
\end{equation}
where the minimum is taken over all possible pure state
decompositions of $\rho_{AB}={\sum_k p_k \ket{\psi_k}_{AB}\bra{\psi_k}}$.

In addition to separability criterion,
CREN is known to be a good entanglement measure in bipartite quantum systems
with the property of entanglement monotone:
CREN does not increase under local operations and classical communication~\cite{LCOK}.
Furthermore, as in the definition of {\em entanglement of formation}~\cite{bdsw},
CREN can also be considered as the amount of entanglement
needed to prepare the state $\rho_{AB}$ quantified by the negativity,
that is, the concept of {\em formation}.

As a dual quantity to CREN, NoA is defined as
\begin{equation}
\mathcal{N}^a(\rho_{AB})=\max \sum_k p_k
\mathcal{N} (\ket{\psi_k}_{AB}), \label{NoA}
\end{equation}
where the maximum is taken over all possible pure-state decompositions of $\rho_{AB}$~\cite{kds}.

For the case when $\rho_{AB}$ is a pure state, its CREN as well as its NoA coincide
with the original negativity, that is,
\begin{equation}
{\mathcal N}_c\left(\ket{\psi}_{AB}\right)={\mathcal N}\left(\ket{\psi}_{AB}\right)
={\mathcal N}^a\left(\ket{\psi}_{AB}\right),
\label{coin}
\end{equation}
for any pure state $\ket{\psi}_{AB}$.

We note that
NoA in Eq.~(\ref{NoA}) is mathematically dual to CREN in Eq.~(\ref{CREN})
because one of them is the minimal average of entanglement over all possible pure-state decompositions
whereas the other is defined as the maximal one.
Moreover, if we introduce a reference system $C$ for a purification of $\rho_{AB}$,
it can be easily shown that there is a one-to-one correspondence
between the set of all possible pure-state decompositions of $\rho_{AB}$
and the set of all possible rank-one measurements on the system $C$.
In other words, ${\mathcal N}^a(\rho_{AB})$ is the maximal entanglement
that can be distributed between the systems $A$ and $B$ with the assistance of the environment $C$,
whereas ${\mathcal N}_c(\rho_{AB})$ is the minimal amount of entanglement
needed to prepare $\rho_{AB}$.
Thus, NoA can also be considered as the quantity physically dual to CREN;
the possible distribution versus the concept of formation.

\subsection{Analytic Lower Bound of the Distributed Entanglement}
\label{Subsec: lbound}

Let us consider two-qudit pure states
$\ket{\phi}_{AB}=\sum_{j=0}^{d-1} \sqrt{p_j}\ket{j}_A\ket{j}_B$ and
$\ket{\psi}_{CD}=\sum_{j'=0}^{d-1} \sqrt{q_{j'}}\ket{j'}_C\ket{j'}_D$
in two bipartite quantum systems $AB$ and $CD$ respectively,
and let
\begin{equation}
\ket{\Psi_{k,l}}=\frac{1}{\sqrt{d}}\sum_{j=0}^{d-1}\omega^{jl}\ket{j,j+k},
\label{dmax}
\end{equation}
be a maximally entangled state in $d\otimes d$ quantum systems
where $\omega=\exp(2\pi i/d)$ is the $d$th root of unity.
(Throughout this paper, all indices are considered to be integers modulo $d$.)
The set $\mathcal{S}\equiv\{\ket{\Psi_{k,l}}: k, l=0,\ldots, d-1\}$ forms an orthonormal basis for
the $d\otimes d$ quantum system, and we furthermore have
\begin{eqnarray}
\frac{1}{\sqrt{d}}\sum_{l=0}^{d-1}\omega^{-lj}\ket{\Psi_{j'-j,l}}
&=& \frac{1}{d}\sum_{l, m=0}^{d-1}\omega^{l(m-j)}\ket{m, m+j'-j}\nonumber\\
&=& \sum_{m=0}^{d-1}\delta_{m,j}\ket{m, m+j'-j}\nonumber\\
&=& \ket{j, j'},
\end{eqnarray}
for each $j,j' \in \{0,\ldots, d-1\}$. Thus it follows that
\begin{eqnarray}
\ket{\Phi}_{ABCD}
&\equiv&
\ket{\phi}_{AB}\otimes\ket{\psi}_{CD}\nonumber\\
&=&\sum_{j=0}^{d-1}\sum_{j'=0}^{d-1}\sqrt{p_j q_{j'}}\ket{j, j'}_{AC}\ket{j, j'}_{BD}\nonumber\\
&=&\frac{1}{\sqrt{d}}\sum_{l=0}^{d-1}
\sum_{j, j'=0}^{d-1}\sqrt{p_j q_{j'}}\omega^{-lj}\ket{j, j'}_{AC}\ket{\Psi_{j'-j,l}}_{BD}.
\nonumber \\
\label{abcd}
\end{eqnarray}

Assume that the system $BD$ is measured in the basis $\mathcal{S}$
and the measurement outcome is $\ket{\Psi_{k,l}}_{BD}$ for some $k$ and $l$.
Then the resulting state in the system $AC$ is
the normalized vector of
\begin{eqnarray}
\ket{\tilde{\psi}_{k,l}}_{AC}
&=&\frac{1}{\sqrt{d}}\sum_{j=0}^{d-1}
\sqrt{p_j q_{j+k}}\omega^{-lj}\ket{j, j+k}_{AC}\nonumber\\
&=&\sqrt{r_{k,l}}\ket{\psi_{k,l}}_{AC},
\end{eqnarray}
where $r_{k,l}$ is the probability of the measurement outcome, that is,
$r_{k,l}=\left|\inn{\tilde{\psi}_{k,l}}{\tilde{\psi}_{k,l}}\right|$.

Because we have
${\mathcal N}\left(|\tilde{\psi}\rangle\right)=p{\mathcal N}\left(\ket{\psi} \right)$
for any unnormalized state $|\tilde{\psi}\rangle=\sqrt{p}\ket{\psi}$ with $\inn{\psi}{\psi}=1$,
the average negativity
that can be distributed on the system $AC$ from the state $\ket{\Phi}_{ABCD}$ in Eq.~(\ref{abcd})
by the joint measurement on the system $BD$ with respect to the basis $\mathcal{S}$ is
\begin{eqnarray}
\sum_{k,l}p_{k,l} \mathcal{N}\left(\ket{\psi_{k,l}}_{AC}\right)
&=&\sum_{k,l}\mathcal{N}\left(\ket{\tilde{\psi}_{k,l}}_{AC}\right)\nonumber\\
&=& \frac{2}{d-1}\sum_{k=0}^{d-1}\sum_{j<j'}\sqrt{p_j q_{j+k}}\sqrt{p_{j'} q_{j'+k}}
\nonumber \\
&=& \frac{2}{d-1}\sum_{k=0}^{d-1}\sum_{j<j'}\sqrt{p_j p_{j'}}\sqrt{q_{j+k} q_{j'+k}}.
\nonumber\\
\label{ave}
\end{eqnarray}
Thus, from Eq.~(\ref{ave}) together with the definition of NoA in Eq.~(\ref{NoA}), we have
\begin{equation}
\mathcal{N}^a \left(\rho_{AC}\right)
\geq \frac{2}{d-1}\sum_{k=0}^{d-1}\sum_{j<j'}\sqrt{p_j p_{j'}}\sqrt{q_{j+k} q_{j'+k}},
\label{eq: bound}
\end{equation}
where $\rho_{AC}=\T_{BD}\ket{\Phi}_{ABCD}\bra{\Phi}$, which is an analytic lower bound
of the possible RDE onto the system $AC$.

Remark that, for arbitrary pair of pure bipartite states,
the bound in~(\ref{eq: bound}) also holds
since local unitary operations preserve the inequality,
and that the bound is a tight one
since it is straightforward to check that the inequality is saturated when both $\ket{\phi}_{AB}$ and
$\ket{\psi}_{CD}$ are maximally entangled states.


\section{Bound on Possible Remotely-Distributed Entanglement}
\label{Sec: bound}
In this section, using the analytic lower bound in~(\ref{eq: bound}),
we show that the possible RDE on the system $AC$ by joint measurement on system $BD$
is not less than the product of two amounts of entanglement for $\ket{\phi}_{AB}$ and $\ket{\psi}_{CD}$
for several cases, by employing CREN and NoA.

\subsection{Low-dimensional systems: Qubits and Qutrits}
\label{Subsec: qubits and qutrits}

We first consider the case that $d=2$.
Then we have $\ket{\phi}_{AB}=\sqrt{p_0}\ket{00}_{AB}+\sqrt{p_1}\ket{11}_{AB}$ and
$\ket{\psi}_{CD}=\sqrt{q_0}\ket{00}_{CD}+\sqrt{q_1}\ket{11}_{CD}$.
In this case, the right-hand side of the inequality~(\ref{eq: bound}) becomes
\begin{equation}
4\sqrt{p_0p_1}\sqrt{q_0q_1}
=\mathcal{N}_c\left(\ket{\phi}_{AB}\right)\mathcal{N}_c\left(\ket{\psi}_{CD}\right),
\label{2rhs}
\end{equation}
and we have
\begin{equation}
\mathcal{N}^a \left(\rho_{AC}\right)
\geq \mathcal{N}_c\left(\ket{\phi}_{AB}\right)\mathcal{N}_c\left(\ket{\psi}_{CD}\right).
\label{2bound}
\end{equation}

We now take account of the case that $d=3$.
Then we have $\ket{\phi}_{AB}=\sqrt{p_0}\ket{00}_{AB}+\sqrt{p_1}\ket{11}_{AB}+\sqrt{p_2}\ket{22}_{AB}$ and
$\ket{\psi}_{CD}=\sqrt{q_0}\ket{00}_{CD}+\sqrt{q_1}\ket{11}_{CD}+\sqrt{q_2}\ket{22}_{CD}$.
As in the case that $d=2$, the right-hand side of the inequality~(\ref{eq: bound}) becomes
\begin{equation}
\left(\sqrt{p_0p_1}+\sqrt{p_0p_2}+\sqrt{p_1p_2}\right)\left(\sqrt{q_0q_1}+\sqrt{q_0q_2}+\sqrt{q_1q_2}\right),
\label{3rhs}
\end{equation}
which is equal to $\mathcal{N}_c\left(\ket{\phi}_{AB}\right)\mathcal{N}_c\left(\ket{\psi}_{CD}\right)$.
Therefore, we are ready to have the following theorem.

\begin{Thm}\label{Thm: 23}
For any states $\ket{\phi}_{AB}$ and $\ket{\psi}_{CD}$
in $d\otimes d$ quantum systems $AB$ and $CD$  with $d=2, 3$,
the possible RDE onto the system $AC$
by joint measurement of the systems $B$ and $D$
is always bounded below by the product of two CREN values
of $\ket{\phi}_{AB}$ and $\ket{\psi}_{CD}$, that is,
\begin{equation}
\mathcal{N}^a \left(\rho_{AC}\right)
\geq \mathcal{N}_c\left(\ket{\phi}_{AB}\right)\mathcal{N}_c\left(\ket{\psi}_{CD}\right).
\label{eq:dbound}
\end{equation}
\end{Thm}

\subsection{General Quantum Systems}
\label{Subsec: qudits}

While there is a simple equality between the right-hand side of the inequality~(\ref{eq: bound})
and $\mathcal{N}_c\left(\ket{\phi}_{AB}\right)\mathcal{N}_c\left(\ket{\psi}_{CD}\right)$
in low-dimensional quantum systems,
it can be easily checked that such a direct equality does not hold
for general case of higher-dimensional systems when $d\geq4$.
Here, we provide two sufficient conditions
for general states of $\ket{\phi}_{AB}$ and $\ket{\psi}_{CD}$
to have the same relation as in the inequality~(\ref{eq:dbound}).

One simple sufficient condition is that
$\ket{\phi}_{AB}$ or $\ket{\psi}_{CD}$ is a $d$-dimensional maximally entangled state.
Suppose $\ket{\phi}_{AB}$ is a maximally entangled state, that is,
\begin{equation}
\ket{\phi}_{AB}=\frac{1}{\sqrt{d}}\sum_{j=0}^{d-1}\ket{jj}_{AB}.
\label{dmax}
\end{equation}
Because ${\mathcal N}_c\left(\ket{\phi}_{AB} \right)={\mathcal N}\left(\ket{\phi}_{AB} \right)=1$,
it can be easily checked that the right-hand side of the inequality~(\ref{eq: bound})
becomes
\begin{eqnarray}
\frac{2}{d-1}\sum_{j<j'}\sqrt{q_{j} q_{j'}}
&=&{\mathcal N}_c\left(\ket{\psi}_{AB} \right)\nonumber\\
&=&{\mathcal N}_c\left(\ket{\phi}_{AB} \right){\mathcal N}_c\left(\ket{\psi}_{AB} \right).
\label{dmaxrhd}
\end{eqnarray}
Hence, in this case, we can readily obtain the same inequality as in (\ref{eq:dbound}).

Now, let us consider another sufficient condition
that the states $\ket{\psi}_{AB}$ and $\ket{\phi}_{CD}$ have the same Schmidt coefficients,
that is,
$p_j = q_j$ for each $j=0,\ldots, d-1$.

First assume that $d\ge 3$ is odd.
Then there exist a positive integer $n$ such that $d=2n+1$.
Let us define a set $S$ as
\begin{equation}
S\equiv\{ \sqrt{p_ip_j}: 0\leq i < j \leq d-1\},
\label{S}
\end{equation}
and, for each $l=1,\ldots, n$, its subset $P_l$ as
\begin{equation}
P_l\equiv\{\sqrt{p_0p_l}, \sqrt{p_1p_{1+l}}, \ldots, \sqrt{p_{2n}p_{2n+l}}\},
\end{equation}
where all indices are integers modulo $d$ as mentioned before.
It is then straightforward to check that
the subsets $P_l$ and $P_{l'}$ do not intersect with each other if $l\neq l'$,
and $S=\bigcup_{l=1}^{n}P_l$.
In other words, $P_l$'s form a partition of the set $S$.

Now, for each $l=1, \ldots, n$, let
\begin{equation}
s_l\equiv\sqrt{p_0p_l}+\sqrt{p_1p_{1+l}}+\cdots +\sqrt{p_{2n}p_{2n+l}},
\label{sl1}
\end{equation}
which is the sum of all elements in $P_l$.
Then the right-hand side of the inequality~(\ref{eq: bound}) becomes
\begin{eqnarray}
\frac{2}{d-1}\sum_{k=0}^{d-1}\sum_{j<j'}\sqrt{p_j p_{j'}}\sqrt{p_{j+k} p_{j'+k}}
=\frac{1}{n}\sum_{l=1}^{n}s_l^2.
\label{rhsodd}
\end{eqnarray}
Furthermore, we obtain
\begin{equation}
\mathcal{N}_c\left(\ket{\phi}_{AB}\right)\mathcal{N}_c\left(\ket{\psi}_{CD}\right)
=\left(\mathcal{N}_c\left(\ket{\phi}_{AB}\right)\right)^2
=\left(\frac{1}{n}\sum_{l=1}^{n}s_l\right)^2.
\label{compodd}
\end{equation}
By letting $\mathcal{K}=\sum_{l=1}^{n}s_l^2/n$
and $\mathcal{L}=\left(\sum_{l=1}^{n}s_l/n\right)^2$,
we have the following equalities
\begin{equation}
n^2\left(\mathcal{K}-\mathcal{L}\right)=(n-1)\sum_{l=1}^{n}s_l^2-2\sum_{l<l'}s_ls_{l'}
=\sum_{l<l'}\left(s_l-s_{l'}\right)^2,
\label{diffodd}
\end{equation}
which is clearly nonnegative.
Hence, it follows that
\begin{equation}
\mathcal{N}^a \left(\rho_{AC}\right)\geq \mathcal{K} \geq \mathcal{L}
=\mathcal{N}_c\left(\ket{\phi}_{AB}\right)\mathcal{N}_c\left(\ket{\psi}_{CD}\right).
\label{oddbound}
\end{equation}

We now assume that $d$ is even such that $d=2m$ for some positive integer $m$.
Similarly,
for each $l=1,\ldots, m-1$,
let us define the subset $Q_l$ of the set $S$ in Eq.~(\ref{S}) as
\begin{equation}
Q_l\equiv\{\sqrt{p_0p_l}, \sqrt{p_1p_{1+l}} \ldots, \sqrt{p_{2m-1}p_{2m-1+l}}\}
\end{equation}
and the subset $Q_m$ as
\begin{equation}
Q_m\equiv\{\sqrt{p_0p_m}, \sqrt{p_1p_{1+m}} \ldots, \sqrt{p_{m-1}p_{2m-1}} \}.
\end{equation}
Then it is also straightforward to check that
$Q_l\cap Q_{l'}$ is the empty set for $l\neq l'$ and $S=\bigcup_{l=1}^{m}Q_l$.
Let us define
\begin{equation}
t_l\equiv\sqrt{p_0p_l}+\sqrt{p_1p_{1+l}}+\cdots +\sqrt{p_{2m-1}p_{2m-1+l}},
\label{sleven}
\end{equation}
for each $l=1, \ldots, m-1$, and
\begin{equation}
t_m\equiv\sqrt{p_0p_m}+\sqrt{p_1p_{1+m}}+\cdots +\sqrt{p_{m-1}p_{2m-1}},
\label{sleven2}
\end{equation}
which are the sums of all elements in $Q_l$ for $l=1, \ldots, m-1$ and $Q_m$, respectively.

Now, the right-hand side of the inequality~(\ref{eq: bound}) becomes
\begin{eqnarray}
\frac{2}{d-1}\sum_{k=0}^{d-1}\sum_{j<j'}\sqrt{p_j p_{j'}}\sqrt{p_{j+k} p_{j'+k}}
=\frac{2}{2m-1}\left(\sum_{l=1}^{m-1}t_l^2+2t_m^2\right).
\nonumber\\
\label{rhseven}
\end{eqnarray}
Moreover, we have
\begin{eqnarray}
\mathcal{N}_c\left(\ket{\phi}_{AB}\right)\mathcal{N}_c\left(\ket{\psi}_{CD}\right)
&=&\left(\mathcal{N}_c\left(\ket{\phi}_{AB}\right)\right)^2\nonumber\\
&=&\left(\frac{2}{2m-1}\right)^2\left(\sum_{l=1}^{m-1}t_l+t_m\right)^2.\nonumber\\
\label{compeven}
\end{eqnarray}
By letting $\mathcal{U}={2}\left(\sum_{l=1}^{m-1}t_l^2+2t_m^2\right)/\left({2m-1}\right)$ and
$\mathcal{V}=4\left(\sum_{l=1}^{m-1}t_l+t_m\right)^2/\left({2m-1}\right)^2$,
we obtain
\begin{widetext}
\begin{eqnarray}
\frac{\left(2m-1\right)^2}{2}\left(\mathcal{U}-\mathcal{V}\right)
&=&(2m-1)\left(\sum_{l=1}^{m-1}t_l^2+2t_m^2\right)-2\left(\sum_{l=1}^{m-1}t_l+t_m\right)^2 \nonumber\\
&=&(2m-1)\sum_{l=1}^{m-1}t_l^2+2(2m-1)t_m^2
-2\left[\left(\sum_{l=1}^{m-1}t_l\right)^2 -t_m^2-2t_m\sum_{l=1}^{m-1}t_l\right]\nonumber\\
&=&2(m-2)\sum_{l=1}^{m-1}t_l^2-4\sum_{l<l'}t_lt_{l'}
+\sum_{l=1}^{m-1}t_l^2 -4t_m\sum_{l=1}^{m-1}t_l+4(m-1)t_m^2\nonumber\\
&=&2\sum_{l<l'}\left(t_l-t_{l'}\right)^2+\sum_{l=1}^{m-1}\left(t_l-2t_m\right)^2,
\label{diffeven}
\end{eqnarray}
\end{widetext}
which is clearly nonnegative as well.
Therefore, it follows that
\begin{equation}
\mathcal{N}^a \left(\rho_{AC}\right)\geq \mathcal{U} \geq \mathcal{V}
=\mathcal{N}_c\left(\ket{\phi}_{AB}\right)\mathcal{N}_c\left(\ket{\psi}_{CD}\right).
\label{oddbound}
\end{equation}

Now, we are ready to have the following theorem.
\begin{Thm}\label{Thm: dbound}
Let $\ket{\phi}_{AB}$ and $\ket{\psi}_{CD}$ be any pure states
in two $d\otimes d$ quantum systems $AB$ and $CD$ respectively,
and assume that they have the same Schmidt coefficients or one of them is maximally entangled.
Then the possible RDE onto the system $AC$ by joint measurement on the systems $B$ and $D$
is always bounded below by the product of two CREN values for $\ket{\phi}_{AB}$ and $\ket{\psi}_{CD}$,
that is,
\begin{equation}
\mathcal{N}^a \left(\rho_{AC}\right)
\geq \mathcal{N}_c\left(\ket{\phi}_{AB}\right)\mathcal{N}_c\left(\ket{\psi}_{CD}\right).
\label{eq: dbound}
\end{equation}
\end{Thm}

Due to the continuity of the inequality~(\ref{eq: dbound})
with respect to the Schmidt coefficients of $\ket{\phi}_{AB}$ and $\ket{\psi}_{CD}$,
we note here that Theorem~\ref{Thm: dbound} also holds
if $\ket{\phi}_{AB}$ and $\ket{\psi}_{CD}$ have similar Schmidt coefficients to each other, or
one of them is nearly maximally entangled.

\section{Summary}\label{Sec: Conclusions}

We have provided the bound for RDE using CREN and NoA.
For a pair of pure states $\ket{\phi}_{AB}$ and $\ket{\psi}_{CD}$
in quantum systems $AB$ and $CD$ respectively,
we have shown that the possible RDE on $AC$ by joint measurement of $BD$,
${\mathcal N}^a\left(\rho_{AC} \right)$,
is bounded below by the product of two amounts of entanglement
in terms of CREN or the negativity,
$\mathcal{N}_c\left(\ket{\phi}_{AB}\right)\mathcal{N}_c\left(\ket{\psi}_{CD}\right)$,
for the case of low-dimensional quantum systems.
We have also presented some sufficient conditions of the states $\ket{\phi}_{AB}$ and $\ket{\psi}_{CD}$,
for which the result of low-dimensional systems
can be generalized into higher-dimensional quantum systems.
Our result also provides an operational interpretation of NoA
as the capacity of possible RDE.

JSK was supported by {\it i}CORE, MITACS, and USARO,
and S.L. was supported by Basic Science Research Program
through the National Research Foundation of Korea (NRF)
funded by the Ministry of Education, Science and Technology (Grant No.~2009-0076578).


\end{document}